\documentclass{article}
\usepackage[T1]{fontenc}

\usepackage{microtype}
\usepackage{graphicx}
\usepackage{subcaption}
\usepackage{booktabs} 
\usepackage{lipsum}
\usepackage{xcolor} 
\usepackage{hyperref}
\usepackage{url}
\usepackage{multirow}
\usepackage{glossaries}
\usepackage{graphicx}
\usepackage{booktabs}
\usepackage{pifont}
\usepackage[table]{xcolor}
\usepackage{wrapfig}
\usepackage[most]{tcolorbox}

\setacronymstyle{long-short}
\newacronym{ssl}{SSL}{self-supervised learning}
\newacronym{sota}{SOTA}{state-of-the-art}
\newacronym{mae}{MAE}{masked autoencoder}
\newacronym{vit}{ViT}{vision transformer}
\newacronym{mim}{MIM}{masked image modeling}
\newacronym{ste}{STE}{straight-through-estimator}

\newcommand{\cmark}{\ding{51}}
\newcommand{\xmark}{\ding{55}}

\usepackage[accepted]{icml2026}

\usepackage{amsmath}
\usepackage{amssymb}
\usepackage{mathtools}
\usepackage{amsthm}
\usepackage{enumitem}
\newtcolorbox{mybox}[1][]{
        enhanced,
        title=#1,
        coltitle=black,
        left=4pt,
        right=4pt,
        top=4.5pt,
        bottom=0pt,
        attach boxed title to top center={yshift=-5pt},
        boxed title style={frame hidden, size=small, colback=lightgray!20},
        sharp corners,
        rounded corners,
        arc=7pt,
}

\DeclareMathOperator*{\minimize}{minimize}

\usepackage[capitalize,noabbrev]{cleveref}

\theoremstyle{plain}

\theoremstyle{definition}

\theoremstyle{remark}

\usepackage[textsize=tiny]{todonotes}
\definecolor{posgreen}{RGB}{34, 139, 34}
\definecolor{negred}{RGB}{178, 34, 34}

\icmltitlerunning{Better Audio Transformer Guided by Convex Gated Probing}

\begin{document}

\twocolumn[
  \icmltitle{BAT: Better Audio Transformer Guided by Convex Gated Probing}


  \icmlsetsymbol{equal}{*}

  \begin{icmlauthorlist}
    \icmlauthor{Houtan Ghaffari}{equal,ghent}
    \icmlauthor{Lukas Rauch}{equal,kassel,esp}
    \icmlauthor{Christoph Scholz}{kassel,fraunhofer}
    \icmlauthor{Paul Devos}{ghent}
  \end{icmlauthorlist}

  \icmlaffiliation{ghent}{Ghent University}
  \icmlaffiliation{fraunhofer}{Fraunhofer IEE}
  \icmlaffiliation{kassel}{University of Kassel}
  \icmlaffiliation{esp}{Now at Earth Species Project}

  \icmlcorrespondingauthor{Houtan Ghaffari}{houtan.ghaffari@ugent.be}
  \icmlcorrespondingauthor{Lukas Rauch}{l.rauch@uni-kassel.de}

  \icmlkeywords{SSL, Audio, Probing, Frozen}

  \vskip 0.3in
]



\printAffiliationsAndNotice{\icmlEqualContribution}

\begin{abstract}
    Probing is widely adopted in computer vision to faithfully evaluate self-supervised learning (SSL) embeddings, as finetuning may misrepresent their inherent quality. In contrast, audio SSL models still rely on finetuning because simple probing fails to unlock their full potential and alters their rankings when competing on AudioSet. Hence, a robust and efficient probing mechanism is required to guide the trajectory of audio SSL towards reliable and reproducible methods. We introduce \emph{Convex Gated Probing} (CGP), a prototype-based method that significantly closes the gap between finetuning and probing in audio. CGP efficiently utilizes all frozen layers via a gating mechanism and exposes the location of latent task-relevant information. Guided by CGP as a reliable post-hoc evaluation probe, we rework the entire SSL pipeline of current best performing audio models that use legacy implementations of prior SSL methods. By refining data preprocessing, model architecture, and pretraining recipe, we introduce \emph{Better Audio Transformer} (BAT), and establish new SOTA on audio benchmarks.
\end{abstract}

\section{Introduction}
Self-supervised learning (SSL) has become the foundation of modern deep learning, achieving state-of-the-art (SOTA) performance across modalities~\citep{he2022_maskedauto,chen2020simple,baevski2022_data2vec2}. In audio, progress has largely been achieved by adapting vision-based methods from images to spectrograms~\citep{huang2022_AMAE}. While prior results on the AudioSet~\citep{gemmeke2017_audioset} benchmark are surpassed by novel SSL models~\citep{chen2024_EAT,alex2025_SSLAM}, the evaluation methodology remains underdeveloped~\citep{rauch2025unmute}. Although benchmarks such as SUPERB~\citep{yang21_superb} and HEAR~\citep{HEARbenchmark22} utilize frozen evaluation, the current pursuit of SOTA performance on AudioSet relies on finetuning. While finetuning may deliver the highest downstream performance, it introduces confounding variables (e.g., hyperparameter sensitivity) that can obscure true progress~\citep{kumar2022FTdistortspretrainedfeatures}. As we show in this work, current SOTA results may reflect better optimization procedures rather than superior SSL representations. Although frozen-feature probing has become an important evaluation technique in computer vision~\citep{oquab23_DINOv2,darcet2025_capi}, simple probes fail to unlock the potential of audio embeddings, leading to a performance gap that falsely justifies reliance on finetuning~\citep{rauch2025unmute}. Reproducibility remains an additional challenge for recent SSL models in audio. For instance, EAT~\citep{chen2024_EAT} and SSLAM~\citep{alex2025_SSLAM} are built upon legacy code via fairseq~\citep{otto19_fairseq} and inherit methodologies from Data2Vec 2.0~\citep{baevski2022_data2vec2} and Audio-MAE~\citep{huang2022_AMAE}. These implementations contain undocumented architectural and optimization details, which complicate reproducibility.

This work systematically improves the recent audio SSL models~\citep{baevski2022_data2vec2,chen2024_EAT,alex2025_SSLAM} and prototype-based probing methods~\citep{rauch2025unmute}, resulting in the following contributions:

\begin{mybox}[\textbf{Contributions}]
\begin{enumerate}[leftmargin=0.25cm]
\item \textbf{Convex Gated Probing (CGP):} A SOTA probing protocol for reproducible and faithful evaluation of audio SSL models. CGP significantly closes the performance gap between probing and finetuning.
\item \textbf{Better Audio Transformer (BAT):} A modernized SSL model that achieves SOTA performance on audio benchmarks. BAT includes a refined audio preprocessing pipeline, improves the ViT's attention module, and consequently enhances the quality of the SSL targets for pretraining.
\item \textbf{Standardized methodology:} We identify inconsistencies in SOTA audio pipelines and provide transparent, extended, and reproducible implementations for BAT and also prior SOTA models. The code is available at: \url{https://github.com/houtan-ghaffari/BAT_ICML2026}
\end{enumerate}
\end{mybox}

\begin{figure*}[!t]
\centering
\includegraphics[width=0.95\textwidth]{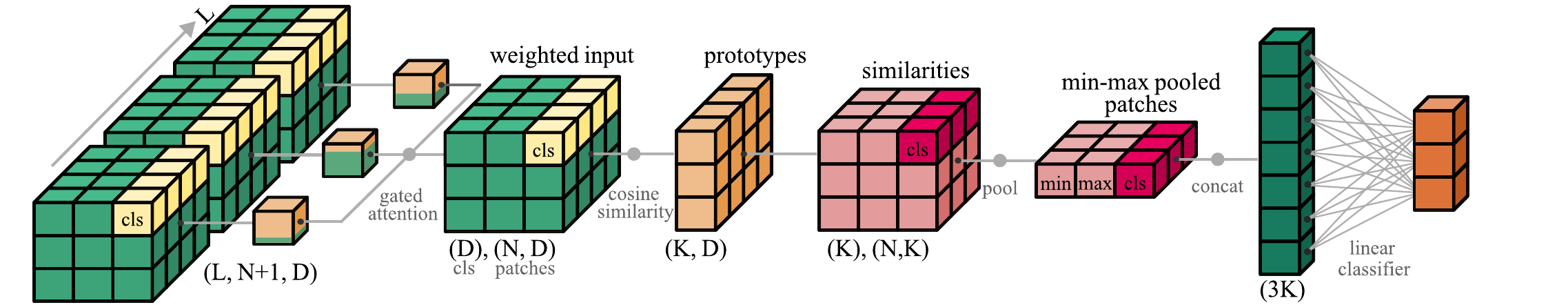}
\caption{\textbf{Convex Gated Probing (CGP) method.} We illustrate the probing process of a spectrogram embedding for a ViT backbone. CGP applies a learnable soft-gating vector (softmax) to compute a weighted sum of embeddings from all layers ($L$). The gating aggregates the hierarchy into a single representation, which is then compared against $K$ prototypes. The cosine similarities of the patch embeddings are min-max pooled and concatenated with the ones from the \texttt{cls}-token, resulting in $3K$ features for a linear classifier.}
\label{fig:protofull}
\end{figure*}

\section{Related Work}
\subsection{Probing}

\textbf{Evaluation with frozen embeddings.} While audio benchmarks~\citep{HEARbenchmark22,yang21_superb} use standardized probing protocols, the pursuit of SOTA performance on AudioSet~\citep{gemmeke2017_audioset} relies predominantly on end-to-end finetuning~\citep{rauch2025unmute,rauch2024deepactivelearningavian}. Although finetuning might maximize model performance, it can obscure the intrinsic quality of the representation by overwriting it~\citep{kumar2022FTdistortspretrainedfeatures}. Conversely, standard linear probes often underestimate embeddings, particularly in \gls*{mim}, since the semantic information is dispersed across token maps and layers rather than concentrated in the final \texttt{cls}-token~\citep{2025beyondcls,alkin2025mim}. While attentive pooling~\citep{darcet2025_capi,psomas2026attentionplease} improves an embedding's summary, it forces a single-vector description. Recent work in audio shifts toward multi-vector aggregation. \citet{niizumi2022_linear_pre} preserve the structure of patch tokens on the frequency axis by averaging the temporal axis of the embeddings, while prototypical probes~\citep{rauch2025canmasked, rauch2025unmute} learn class-wise prototypes directly from the patch token map. By disentangling spatially dispersed events, prototypical probing shows that the large gap between frozen embeddings and finetuned models in audio is an artifact of the pooling method, positioning prototypical probing as a competitive alternative for SOTA evaluation.

\textbf{Layer-aware evaluation.} Previous works address the spatial bottleneck of the embeddings. However, extracting SSL embeddings from the last layer does not necessarily preserve intermediate information that may be better suited for a downstream task~\citep{lee2023surgicalFT,evci22_Head2Toe}. This is particularly evident in \gls*{mim} architectures, where lightweight decoders might force the final encoder layers to assist in low-level reconstruction, causing semantic information to peak in middle layers~\citep{he2022_maskedauto,alkin2025mim}. Thus, it requires a supervised adaptation step to concentrate information in the final layer's \texttt{cls}-token~\citep{rauch2025canmasked}. Recent works in vision introduce alternative layer-aware strategies that utilize all available layers to adapt the model to a downstream task. Head2Toe~\citep{evci22_Head2Toe} first concatenates the embedding of all layers and employs a group lasso regularization to select the most informative features, which may require significant memory and computation during the feature selection phase. Side-Tuning~\citep{sax2020sidetuning} trains a lightweight network in parallel, and sums its weights with the frozen backbone weights. More recently, Visual Query Tuning (VQT)~\citep{tu2023vqt} introduces per-layer learnable query tokens into the encoder to summarize intermediate features via attention, and concatenates them to capture the dispersed semantics of a frozen backbone. Additionally, H2T-DFR~\citep{wajid2024notonly} combines Head2Toe with deep feature reweighting to combat spurious correlations. 

\textbf{Position of this paper in probing.} We propose CGP, a layer-aware probing method as a faithful alternative to exhaustive finetuning. This method extends and improves the binarized prototypes of \citet{rauch2025canmasked} and also resolves the hierarchical information bottleneck by accessing the full depth of the backbone via a learnable soft-gating mechanism. Unlike VQT~\citep{tu2023vqt}, CGP operates outside the architecture, avoiding internal modifications or attention bias introduced by the SSL objective. To prevent high feature dimensions as in Head2Toe, we project the features into a prototype space and apply pooling to the prototype activations. Also, CGP utilizes both patch tokens and the \texttt{cls}-token (if available) to maximize information extraction and disentangling from SSL models.

\subsection{Self-Supervised Learning}
\textbf{Negative-free contrastive learning.} BYOL~\citep{grill2020_byol} was a breakthrough in SSL by demonstrating that contrastive learning can succeed without negative pairs. It directly pushes the embeddings of two positive views closer together using a Siamese design~\citep{NIPS1993_288cc0ff,chen2020_simsiam}. Although this task admits trivial solutions, BYOL prevents this by: (i) incorporating the teacher-student framework~\citep{bucilua2006model,tarvainen2017mean,hinton2015distilling,lillicrap2019continuouscontroldeepreinforcement,he2020momentum}, where the teacher (target model) is the exponential moving average (EMA) of the student (online model), and (ii) using architectural asymmetry via a prediction head on top of the student. Although SimSiam~\citep{chen2020_simsiam} shows that EMA is not necessary to prevent trivial solutions in BYOL, it is an integral component in modern SSL approaches to achieve top results~\citep{caron2021emerging,baevski2022_data2vec2}.

\textbf{Masked Latent Regression (MLR).} \citet{baevski2022data2vec} propose Data2Vec (D2V) and extend BYOL to a structural MLR task across modalities, rather than learning modality-specific augmentation-invariant embeddings. Following MAE~\citep{he2022_maskedauto}, which in turn was inspired by BERT~\citep{devlin2019bert}, it leverages a Vision Transformer (ViT)~\citep{dosovitskiy2021VIT} and applies intense masking for continuous data modalities. However, it does not drop the masked tokens from the encoder’s inputs and does not use a decoder~\citep{he2022_maskedauto} or architectural asymmetry~\citep{grill2020_byol}. Masking dominates recent SSL algorithms in vision and audio, either as a prediction task~\citep{zhou2022_ibot,baevski2022data2vec,baevski2022_data2vec2}, or an augmentation~\citep{assran2022masked}. Hence, different masking strategies are explored, as they significantly affect the quality of the pretrained embedding. Notably, D2V~\citep{baevski2022data2vec} uses block-masking of BEIT~\citep{bao2022_beit} for the image modality, rather than random masking as in MAE~\citep{he2022_maskedauto}. Masking has also been shown to be an effective data augmentation for clustering-based SSL~\citep{zhou2022_ibot}, even without explicit patch token prediction~\citep{assran2022masked}.

\textbf{Collapse in MLR.} \citet{baevski2022data2vec} notes that preventing collapse in MLR is challenging for continuous data due to the high correlation among neighboring tokens, particularly in audio. Although there are practical SSL algorithms based solely on whitening and feature diversity~\citep{zbontar2021barlow,bardes2021vicreg,ermolov2021whitening}, D2V~\citep{baevski2022data2vec} uses a hyperparameter-free approach that promotes variance by normalizing target representations across the sequence and feature dimensions. Interestingly, BYOL collapses without architectural asymmetry, but masked prediction and target normalization could prevent collapse in D2V. Additionally, \citet{baevski2022data2vec} demonstrate that averaging embeddings from multiple teacher layers yields better regression targets.

\textbf{Improved MLR.} \citet{baevski2022_data2vec2} propose Data2Vec 2.0 (D2V2) by incorporating the efficiency technique from MAE~\citep{he2022_maskedauto} to avoid processing masked tokens in the encoder. They incorporate a lightweight CNN decoder into the student branch to predict the missing masked tokens in the latent space. For image modality, they find it beneficial to add a global loss using the \texttt{cls}-token~\citep{peng2022beit} alongside the local loss for patch token prediction. They also propose inverse-block masking for better contextual representation learning. Additionally, they leverage a multi-masking strategy~\citep{assran2022masked} to reuse the target representations for multiple masked versions of the input in each forward pass.

\textbf{Audio self-supervised models}. There are numerous audio SSL models, which are primarily extensions or direct applications of vision SSL models~\citep{niizumi2021byol,huang2022_AMAE,chen2024_EAT,alex2025_SSLAM,ghaffari2025data}. SSAST~\citep{gong2022ssast} is an early work that introduces ViT~\citep{dosovitskiy2021VIT} to audio tasks. It combines masked spectrogram reconstruction and contrastive learning, although the latter is a reformulation of the former rather than a standard sample-wise contrastive learning~\citep{chen2020simple}. SSAST was introduced for audio prior to MAE~\citep{he2022_maskedauto} for images, but it does not have the efficiency of MAE. \citet{huang2022_AMAE} propose Audio-MAE, an application of MAE to audio spectrograms, which are akin to grayscale images (albeit superficially in terms of 2D structure). Audio-MAE achieved top performance across six audio and speech classification tasks, including AudioSet, the primary benchmark for ranking audio SSL models. Following Audio-MAE, BEATs~\citep{BEATSchen2023} introduce an iterative tokenizer to provide semantic targets for \gls*{mim}, which improves representation quality on AudioSet. Subsequently, \citet{chen2024_EAT} present EAT, a direct application of D2V2~\citep{baevski2022_data2vec2} combined with the spectrogram preprocessing used in Audio-MAE, achieving significant improvements on audio benchmarks. Finally, \citet{alex2025_SSLAM} introduce SSLAM. It starts with pretrained weights from EAT and applies the same algorithm in a second round of pretraining by adding an extra source retention loss to the objective. The source retention task is to predict unmixed targets from partially mixed samples in artificially mixed regions.

\textbf{Position of this paper in SSL.} Per reports on AudioSet, SSLAM is the current SOTA model and surpasses EAT. Our investigation suggests that current SOTA results are not fully reproducible. We observe undocumented implementation details, such as a high loss-scaling heuristic (i.e., a factor of  $8 \times 10^4$ for the global loss relative to the local loss). These two models rely on legacy implementations of D2V2 via fairseq~\citep{otto19_fairseq} and Audio-MAE. We find that finetuning on AudioSet is sensitive to hyperparameter configurations, which complicates the reproduction of current SOTA results. To address these limitations, we systematically modernize the D2V2 framework to develop a better audio transformer, BAT. Rather than relying on finetuning results as an unclear justification for the methodology, we leverage CGP evaluation at every step of the design. First, we integrate a modernized and dataset-independent spectrogram preprocessing pipeline. Then, we introduce gated attention to the audio ViT, which not only improves the baseline but also enhances the SSL targets after rectifying the D2V2 target-generation heuristic. Finally, we increase the decoder's capacity and establish a novel SOTA SSL model. Regarding the evaluation benchmark, we extend prior work by incorporating speech transcription, out-of-distribution generalization, and sound event detection alongside the previous SOTA models' classification benchmarks. We also provide implementations and evaluations for their models.

\section{Audio Masked Latent Regression}\label{sec:mlr}
We first explain the SSL algorithm from D2V2~\citep{baevski2022_data2vec2}, adopted by current SOTA audio models, EAT~\citep{chen2024_EAT}, and SSLAM~\citep{alex2025_SSLAM}.

Denote a ViT by $f_\theta$, $\theta$ being the parameters that we optimize via gradient descent. We refer to this as the online model. We denote an EMA version of the model by $f_{\bar{\theta}}$, with $\bar{\theta}$ at batch step $t$ being updated via $\bar{\theta}^{(t)} = \lambda \bar{\theta}^{(t-1)} + (1 - \lambda) \theta^{(t)}$, i.e., after each batch gradient descent update of the online model. The decay rate, $\lambda$, is either fixed or modified via a linear scheduler. $f_{\bar{\theta}}$ is the target model because it provides the SSL targets for the online model to bootstrap itself, i.e., self-distillation~\citep{caron2021emerging}.

Denote an input audio spectrogram by $x_{spec} \in \mathbb{R}^{T \times F}$, $T$ and $F$ being the number of time and frequency bins, respectively. This input is organized into a sequence of flattened and non-overlapping $k \times k$ patches, denoted by $x \in \mathbb{R}^{N \times k^2}$, where $N = \frac{T}{k} \cdot \frac{F}{k}$. A significant portion of this input, about $80\%$, is randomly masked and removed to get a partial view $x_m \in \mathbb{R}^{n \times k^2}$. We denote the set of masked indices as $\mathcal{I}_m$.

The online model produces $(z_m, o_m) = f_\theta(x_m) \in \mathbb{R}^{n \times D} \times \mathbb{R}^{D}$, which are patch ($z_m$) and cls ($o_m$) tokens embeddings. During the SSL phase, the online model uses a CNN decoder to reconstruct missing patch tokens. Denote it by $g_\phi$, and let $\Tilde{z}_m = g_{\phi}(z_m) \in \mathbb{R}^{N \times D}$ be the online model prediction of the target patch embeddings. We denote the target model outputs by $(z, o) = f_{\bar{\theta}}(x) \in \mathbb{R}^{N \times D} \times \mathbb{R}^{D}$. Although the target and online encoders are identical, their forward passes differ. Let us expand the $l$-th ViT encoder block calculations, 
\begin{align}
    z^{(l)}_a &= \textit{MHSA}(z^{(l-1)}_d),  \\
    z^{(l)}_b &= \textit{LayerNorm}(z^{(l-1)}_d + z^{(l)}_a), \\
    z^{(l)}_c &= \textit{MLP}(z^{(l)}_b),  \\
    z^{(l)}_d &= \textit{LayerNorm}(z^{(l)}_b + z^{(l)}_c),
\end{align}
where $z^{(l)}_d$ is the output, including the \texttt{cls}-token. D2V~\citep{baevski2022data2vec} suggests multiple modifications to create targets, which D2V2~\citep{baevski2022_data2vec2} inherits. First, the target network accumulates the patch embeddings $z^{(l)}_c$ from all layers into a list and drops the \texttt{cls}-tokens. \citet{baevski2022data2vec} find $z^{(l)}_a$ to be uninformative, and $z^{(l)}_c$ is a better target than $z^{(l)}_d$ (we revisit this in Section~\ref{sec:attention_gate}). These targets are standardized across the token axis ($N$). Then, the patch embeddings of all layers are averaged, followed by another standardization along the feature axis ($D$) to produce the target patch embeddings $z$. The normalizations are to prevent collapse and promote variance across tokens and embeddings. Additionally, the target network discards the \texttt{cls}-token embedding and uses $ o = 1/N \sum_j z_j$. The online model is optimized using the following objective,
\begin{align}
\minimize_{\theta, \phi} \ell &= \ell_{global} + \ell_{local}, \\
\ell_{global } &= ||o - o_m||_2^2, \\
\ell_{local} &= \frac{1}{|\mathcal{I}_m|} \sum_{i \in \mathcal{I}_m} ||z_i - \Tilde{z}_{m_i}||_2^2.
\end{align}

\section{Convex Gated Probing}
The primary motivation for CGP is that the best features of an SSL model may not reside in its final layer~\citep{yang21_superb,baevski2020_w2v2}. We illustrate the CGP method in \autoref{fig:protofull}.

\begin{figure}[h!]
\centering
\includegraphics[width=0.95\linewidth]{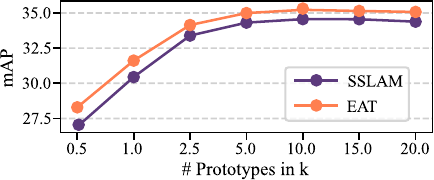}
\caption{\textbf{CGP Ablation on AS-20k.} Increasing the number of prototypes constantly improves results, but yields diminishing returns. Although the best computation-performance trade-off is dataset-dependent, we will use 10k prototypes for all experiments to showcase the efficacy and reliability of CGP.}
\label{fig:cgp_num_prototype_ablation}
\end{figure}

Denote a variant of an already pretrained ViT by $f^{L}_{\theta}$, where $(z, o) = f^L_{\theta}(x) \in \mathbb{R}^{L \times N \times D} \times \mathbb{R}^{L \times D}$ indicate the patch and \texttt{cls}-tokens embeddings of a single input from all layers, with $L$ the number of layers, $N$ the number of patch tokens, and $D$ the size of the embedding. CGP consists of $K$ learnable prototype vectors, denoted by $P \in \mathbb{R}^{K \times D}$. The prototypes and the embeddings are first L2-normalized along the feature dimension,
\begin{equation}
   \hat{P}_k = \frac{P_k}{||P_k||_2}, \quad \hat{z}_{ln} = \frac{z_{ln}}{||z_{ln}||_2}, \quad \hat{o}_l = \frac{o_l}{||o_l||_2}.
\end{equation}
CGP has a learnable weight vector $a \in \mathbb{R}^L$ to aggregate the layers. It is first converted to convex weights via $\alpha = \operatorname{softmax}(a)$, and then,
\begin{equation}
    \bar{z} = \sum_l \alpha_l \hat{z}_l, \quad \bar{o} = \sum_l \alpha_l \hat{o}_l,
\end{equation}
with $\bar{z} \in \mathbb{R}^{N \times D}$ and $\bar{o} \in \mathbb{R}^{D}$. The cosine similarities to prototypes are calculated as, 
\begin{equation}
    s_z = \bar{z} \hat{P}^{\top}, \quad s_o = \bar{o} \hat{P}^{\top},
\end{equation}
with $s_z \in [-1, \, 1]^{N \times K}$ and $s_o \in [-1, \, 1]^{K}$. The patch tokens' similarities, $s_z$, are summarized by taking their maximum and minimum across the tokens, and concatenated with the \texttt{cls}-token similarity,
\begin{equation}
    s = [\min_N s_z, \max_N s_z, s_o],
\end{equation}
resulting in $s \in \mathbb{R}^{3K}$, which is subsequently passed to a linear classifier. \autoref{tab:topaudiossl} presents the results of CGP. We compare them with finetuning and the previous best probing method Protobin~\citep{rauch2025unmute} on EAT and SSLAM, the current SOTA models on AudioSet. Crucially, our finetuning protocol precisely follows the reported procedure in their original works. Despite using publicly available weights, we are unable to replicate the reported SOTA performance via finetuning, highlighting the sensitivity and fragility of finetuning. Consistent with \citet{rauch2025unmute}, our evaluations across ProtoBin, CGP, and finetuning show that SSLAM consistently achieves lower performance than EAT. This suggests that reported improvements in these models may stem from optimization artifacts or dataset differences rather than the models' embeddings. For Protobin and CGP, we use 10k prototypes throughout this work. \autoref{fig:cgp_num_prototype_ablation} shows that this value provides a favorable trade-off, as further increases in the number of prototypes yield diminishing performance gains.

\begin{table}[t!]
    \centering
    \setlength\tabcolsep{0pt}
    \caption{\textbf{Benchmark of EAT and SSLAM on AudioSet}. Comparison between reported performance and our reproduction across finetuning and various probing settings (CGP, Protobin, LP).}
    \label{tab:topaudiossl}
    \small
    \begin{tabular*}{\linewidth}{@{\extracolsep{\fill}} lcccc}
        \toprule
        
        \textbf{Method} & \multicolumn{2}{c}{\textbf{AS-20k}} & \multicolumn{2}{c}{\textbf{AS-2M}} \\
          & \textbf{mAP}     &     \textbf{F1} & \textbf{mAP}    & \textbf{F1} \\
        \midrule
        
        \textcolor{gray!65}{\textit{Reported finetuning}} & & & & \\
        \cmidrule(lr){1-1}
        \textcolor{gray!65}{EAT}~\citep{chen2024_EAT}              & \textcolor{gray!65}{40.2} & \textcolor{gray!65}{-} & \textcolor{gray!65}{48.6} & \textcolor{gray!65}{-} \\
        \textcolor{gray!65}{SSLAM}~\citep{alex2025_SSLAM}          & \textcolor{gray!65}{40.9} & \textcolor{gray!65}{-} & \textcolor{gray!65}{50.2} & \textcolor{gray!65}{-} \\
        \midrule
        
        \textit{Our finetuning} & & & & \\
        \cmidrule(lr){1-1}
        EAT              & \textbf{40.28} & \textbf{30.73} & 47.61 & 36.63 \\
        SSLAM            & 39.97 & 27.03 & \textbf{47.69} & \textbf{36.83} \\
        \midrule
        
        \textit{CGP Probing} & & & & \\
        \cmidrule(lr){1-1}
        EAT              & \textbf{35.20} & \textbf{26.04} & \textbf{41.04} & 19.51 \\
        SSLAM            & 34.62 & 24.58 & 40.20 & \textbf{19.59} \\
        \midrule
        
        \textit{Protobin Probing} & & & & \\
        \cmidrule(lr){1-1}
        EAT              & \textbf{31.22} & \textbf{14.77} & \textbf{32.58} & \textbf{15.0} \\
        SSLAM            & 29.64 & 12.07 & 31.25 & 14.80 \\
        \midrule
        
        \textit{Linear Probing} & & & & \\
        \cmidrule(lr){1-1}
        EAT              & \textbf{17.95} & 15.63 & \textbf{30.36} & \textbf{19.20} \\
        SSLAM            & 17.03 & \textbf{16.86} & 28.92 & 18.44 \\
        \bottomrule
    \end{tabular*}
\end{table}

\section{Ablations for a Better Audio Transformer}
The following ablations gradually introduce and evaluate the methodological enhancements in BAT. The first part is independent of the architecture or SSL, and refines the audio preprocessing pipeline in general. The second part enhances the ViT with an attention gate. Not only does it improve the base model, but it also addresses the problem of uninformative self-attention outputs in D2V~\citep{baevski2022data2vec} and enables us to generate better SSL targets. The last part enhances the decoder in the SSL stage, and by using CGP, we demonstrate that it shifts the semantic features to later layers of the pretrained encoder.

\textbf{Ablation setup.} All pretrainings are conducted exclusively on AS-2M, containing 1,912,024 audio clips of 10 seconds each. The recordings are resampled to 16~kHz. Unlike EAT and SSLAM, we do not rely on the legacy fairseq implementation of D2V2~\citep{baevski2022_data2vec2} and Audio-MAE~\citep{huang2022_AMAE}. We provide a native PyTorch implementation to foster reproducibility. For a fair comparison with EAT and SSLAM, we adopt similar hyperparameters across all pretraining experiments: batch size of 48 with 16 inverse-block masked views per sample, 400~k optimization steps, 50~k steps of linear learning rate warmup from $1\mathrm{e}{-6}$ to $5\mathrm{e}{-4}$ and 350~k steps cosine decay to $1\mathrm{e}{-6}$, and a weight decay of 0.05. In contrast to EAT and SSLAM, we exclude the large loss-scaling factors inherited from the D2V2 framework. We find that scaling the global token loss by a factor of $8\times10^4$ or any other value causes optimization instability and compromises reproducibility. By weighting global and local losses equally, we maintain a simplified and transparent objective that avoids the need for heuristic scaling. We use {bfloat16} mixed-precision optimization instead of {float16}. All ablations are conducted on AS-20k utilizing CGP with 10~k prototypes, 500 steps linear learning rate warmup from $1\mathrm{e}{-6}$ to $1\mathrm{e}{-3}$ and 20~k steps cosine decay to $1\mathrm{e}{-6}$, and a weight decay of 0.05. For these ablations, we use the conventional binary cross-entropy loss.

\subsection{Better Audio Preprocessing Pipeline}
Conventional audio frontends utilize a spectrogram generation, dynamic range compression, and normalization to prepare the input signal. As shown by \citet{ghaffari24_frontends}, frontend choices are critical to performance, influencing feature details, noise sensitivity, and the efficiency of gradient-based optimization. EAT and SSLAM adopt the Audio-MAE frontend: log-compressed mel-spectrogram (filterbanks) and global standardization. This pipeline relies on a legacy implementation that originally concentrated on human speech. Furthermore, using global normalization complicates the practical deployment of pretrained models, as it requires knowledge of downstream dataset statistics.

\begin{table}[ht]
\centering
\caption{\textbf{Audio frontend ablation.} We compare different spectrogram representations and normalization strategies. The \textcolor{gray!70}{baseline} denotes our reproduction of the filterbank inputs used in EAT and AudioMAE, while our \colorbox{posgreen!20}{BAT} configuration is highlighted.}
\label{tab:input-ablation}
\small
\begin{tabular}{lccc}
    \toprule
    \textbf{Data Pipeline} & \textbf{Global Statistics} & \textbf{mAP} & \textbf{F1} \\
    \midrule
    \textcolor{gray!70}{EAT (reproduced)} & \cmark & \textcolor{gray!70}{34.86} & \textcolor{gray!70}{24.28} \\
    \rowcolor{posgreen!20}
    Ours & \xmark & \textbf{35.03} & \textbf{24.75}\\
    \bottomrule
\end{tabular}
\end{table}

\begin{figure*}[h!]
\centering
\includegraphics[width=1.0\textwidth]{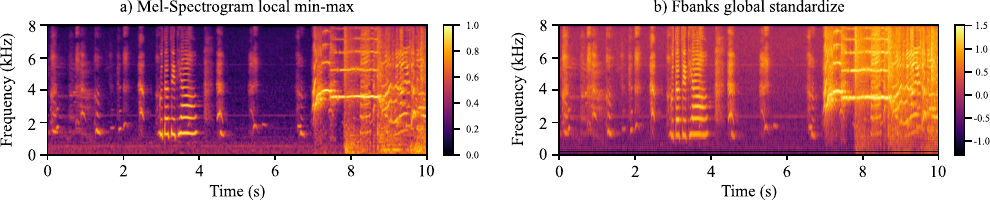}
\caption{\textbf{Impact of audio frontend.} A recording containing the labels [Whimper, Gasp, Speech, Outside, urban or manmade]. (a) Our incorporated audio frontend: Mel-spectrogram with decibel compression and local min-max normalization, exhibiting clear spectral structure and high contrast. (b) Audio-MAE, EAT, and SSLAM: simple log, filtering, Mel-spectrogram, and global standardization. Note the artifacts and blurring, particularly at lower frequencies.}
\label{fig:spectrogram-inputs}
\end{figure*}

We extract mel-spectrograms using a modernized TorchAudio implementation that avoids heuristic filtering and supports efficient batch transformations, ensuring better signal integrity and faster training. We then apply a decibel-scale log compression to improve the dynamic range relative to a log function. Finally, we apply local min-max normalization to scale each mel-spectrogram to $[0, 1]$, thereby effectively suppressing noise and facilitating deployment~\citep{ghaffari24_frontends,ghaffari2025robust}. Results in \autoref{tab:input-ablation} indicate that our pipeline enhances model performance while remaining independent of dataset statistics, ensuring a more robust and flexible solution than legacy implementations. \autoref{fig:spectrogram-inputs} provides a visual comparison of the mel-spectrograms produced by the legacy frontend and our new one.

\subsection{Better Targets with Gated Attention}\label{sec:attention_gate}
\begin{table}[ht]
\centering
\caption{\textbf{Impact of target selection and attention gates.} EOB denotes End-of-Block. The \textcolor{gray!70}{baseline} from the previous section and our proposed \colorbox{posgreen!20}{BAT} configuration are highlighted.}
\label{tab:gated_attention_eob}
\small
\begin{tabular}{l c | l l}
\toprule
\textbf{SSL Target} & \textbf{Gated Attention} & \textbf{mAP} & \textbf{F1}\\
\midrule
\textcolor{gray!70}{MLP} & \xmark & \textcolor{gray!70}{35.03} & \textcolor{gray!70}{24.75} \\
EOB & \xmark & 34.60 & 25.00  \\
MLP & \ding{51} & 35.09 & 26.12 \\
\rowcolor{posgreen!20} EOB & \ding{51} & \textbf{35.42} & \textbf{27.13}  \\
\bottomrule
\end{tabular}
\end{table}

\begin{figure*}[h!]
\centering
\includegraphics[width=1.0\linewidth]{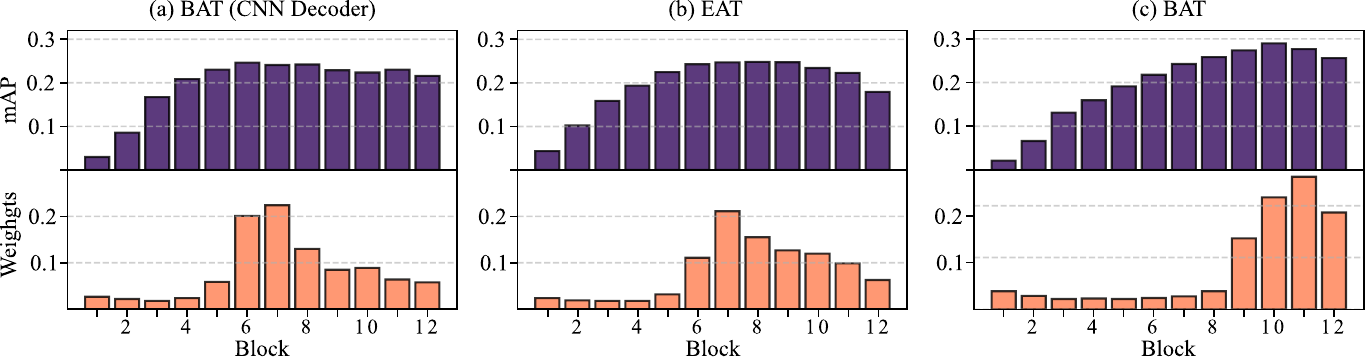}
\caption{\textbf{Layer-wise latent information.} We display the layer-wise latent information quality across three models on AS-20k: (a) BAT with the lightweight CNN (best performer from Table 3), (b) EAT (baseline), and (c) our final BAT (ViT decoder). The top row displays the linear probing performance of each block. The bottom row visualizes the learned gating weights from CGP. Notably, the standard EAT (b) and the CNN-based BAT (a) exhibit a middle-heavy distribution where semantic information peaks early. In contrast, the heavier ViT decoder in the final BAT (c) shifts the semantic peak toward the later layers, improving linear separability at the output.}
\label{fig:cgp_layerwise_info}
\end{figure*}

D2V~\citep{baevski2022data2vec} generates targets by averaging the outputs of the MLP across multiple teacher layers (see Section \ref{sec:mlr}). This design choice emerges from an empirical investigation that shows that using MHSA output as a target leads to representation collapse. Notably, using the End-Of-Block (EOB) output, which sums the MLP and MHSA modules, yields inferior results to MLP. Although their solution improves the results, it violates the semantics of the encoder block as a coherent function. We hypothesize that the performance degradation observed with EOB targets stems directly from the MHSA component, which introduces degenerate inter-token dependencies into the residual connection (e.g., attention sinks), corrupting the semantic quality of the targets. Recent findings in LLMs~\citep{qiu2025gated} demonstrate that applying a sigmoid gate after the attention-weighted value projection improves performance, scaling properties, and training stability. \citet{qiu2025gated} identify two factors contributing to the effectiveness of attention gating in MHSA: (i) introducing non-linearity between value and output projections in the attention block, and (ii) introducing input-dependent sparsity to attention scores, which eliminates attention sinks~\citep{xiao2024efficient} and enhances long-context extrapolation performance. We propose incorporating this gating mechanism~\citep{qiu2025gated} to improve attention and utilizing the EOB output as the SSL target.

We denote the input to MHSA as $x \in \mathbb{R}^{N \times D}$. MHSA applies three linear projections to produce queries, $Q=x W_Q$, keys, $K=x W_K$, and values $V=x W_V$. To accommodate multi-head processing with $H$ heads, these are reshaped and transposed as $\mathbb{R}^{N \times D} \rightarrow \mathbb{R}^{H \times N \times d_h}$, where $d_h = D / H$. The attention-weighted values are,
\begin{equation}
    \bar{V} = \operatorname{softmax}(\frac{QK^{\top}}{\sqrt{d_h}}) V,
\end{equation} 
where $\bar{V} \in \mathbb{R}^{N \times D}$ after transposing and reshaping the results of the above equation (implicitly vectorized on the head axis). The default MHSA applies a final linear projection to produce $O = \bar{V} W_O$. However, the gating mechanism, which we adopt, is as follows, 
\begin{align}\label{eq:attention_gate}
    \Tilde{V} &= \sigma(xW_G) \cdot \bar{V}, \\
    O &= \Tilde{V} W_O,
\end{align} 
where $W_G \in \mathbb{R}^{D \times D}$ and the gate output after the sigmoid activation, $\sigma(\cdot)$, is multiplied element-wise with attention-weighted values. Note that the attention gate is part of the architecture and persists in both the online and target models during pretraining and also downstream probing.

\begin{figure}[ht]
\centering
\includegraphics[width=.99\linewidth]{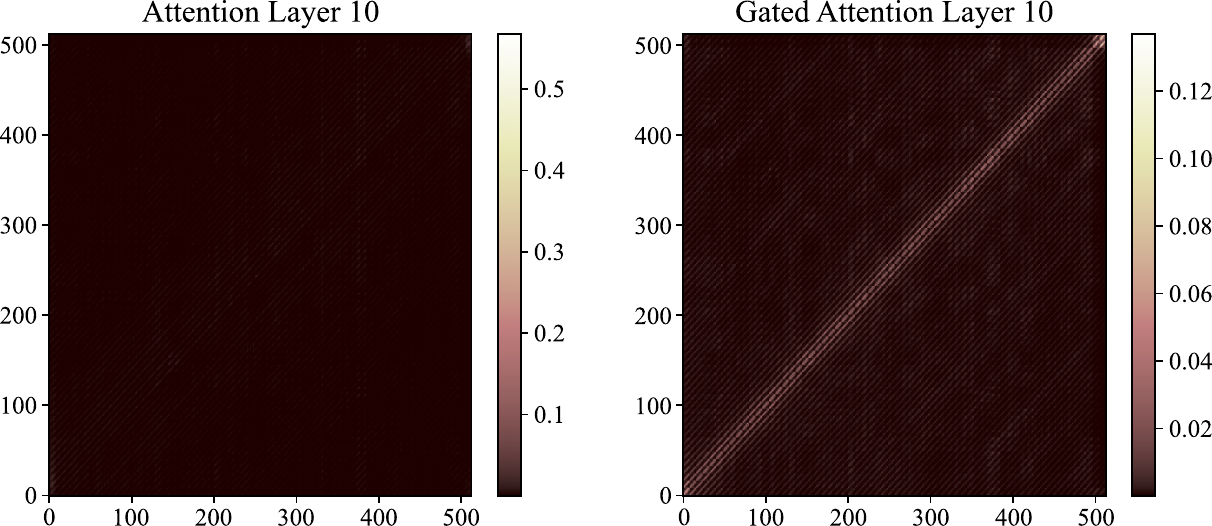}
\caption{\textbf{Impact of gating on attention maps.} Gating distributes attention better and focuses more on the token itself, rather than sinking into one token, primarily the \texttt{cls}-token.}
\label{fig:gated_attention}
\end{figure}

\autoref{tab:gated_attention_eob} presents the ablation of target selection and attention gating. Consistent with prior literature~\citep{baevski2022data2vec}, using EOB outputs as SSL targets without gating degrades downstream performance relative to the baseline that uses MLP outputs as SSL targets (rows 1 and 2). However, integrating the gated attention reverses this observation (rows 3 and 4). Furthermore, notice how the attention gate improves the model, regardless of the SSL targets (compare rows 1 and 3, and rows 2 and 4). Gating not only improves the baseline but also unlocks the potential of the EOB targets, resulting in a coherent forward pass for the target model and the highest performance. This supports our hypothesis that the gating mechanism mitigates the inherent limitations of the MHSA in this SSL dynamic. \autoref{fig:gated_attention} shows the positive effect of gating on attention maps.

\subsection{Better Decoder for a Better Encoder}
D2V2~\citep{baevski2022_data2vec2} uses a lightweight CNN decoder. However, \citet{he2022_maskedauto} show that a sufficiently large decoder improves the encoder performance on downstream tasks. If the decoder is weak, the later layers of the encoder tend to contribute more to masked-token reconstruction than to learning high-level semantics~\citep{huang2022_AMAE}, thereby diminishing the model's effective capacity. This is particularly pronounced in regression-based MIM~\citep{alkin2025mim}, affecting the quality of frozen-feature probing. We replace the six-layer CNN decoder with a six-layer ViT decoder and examine how varying the number of heads and the MLP ratio affects encoder performance.

\begin{table}[ht]
\centering
\caption{\textbf{Impact of decoder.} We replace the lightweight CNN decoder with a ViT. The \textcolor{gray!70}{baseline} from the previous section and our proposed \colorbox{posgreen!20}{BAT} configuration are highlighted.}
\label{tab:decoder}
\small
\begin{tabular}{l c c c | c c}
\toprule
\textbf{Decoder} & \textbf{Depth} & \textbf{Heads} & \textbf{MLP Ratio} & \textbf{mAP} & \textbf{F1}\\
\midrule
CNN & \textcolor{gray!70}{-} & - & \textcolor{gray!70}{-} & \textcolor{gray!70}{35.42} & \textcolor{gray!70}{27.13} \\
ViT & 6 & 6 & 2 & 37.43 & 28.91 \\
\rowcolor{posgreen!20} ViT & 6 & 12 & 4 & \textbf{37.52} & \textbf{29.11} \\
\bottomrule
\end{tabular}%
\end{table}

\autoref{tab:decoder} shows that replacing the CNN with a more expressive ViT decoder yields a performance improvement of 2.0 percentage points (pp) in mAP. Further increasing the decoder capacity leads to 37.52 mAP, a substantial improvement over the reproduced EAT baseline of 34.86 mAP in \autoref{tab:input-ablation}. Additionally, we utilize CGP to analyze the layer-wise distribution of latent information within the frozen encoder. \autoref{fig:cgp_layerwise_info} illustrates this relationship by displaying both the per-block linear probing performance and the learned gating weights of CGP. We observe a strong correlation between these two, suggesting that CGP can automatically identify the most informative layers without the need for exhaustive manual probing of each individual block. Comparing the architectures shows that the CNN-based BAT with attention gates (best performing model from \autoref{tab:gated_attention_eob}) (a) and the standard EAT baseline (b) exhibit a centered distribution, where semantic quality peaks around block 7 and degrades in the final layers. This suggests that the lightweight decoder forces the encoder's deeper layers to retain low-level reconstruction details, even in the latent space. In contrast, our final BAT with a ViT decoder (c) shifts the semantic peak significantly to the right (blocks 10-12). This architectural shift not only improves the final representation but also boosts per-block utility: BAT achieves a peak linear probing accuracy of nearly 30 mAP in the final block, whereas the baseline models struggle to surpass 25 mAP at any depth. This demonstrates that BAT produces richer embeddings that are more linearly separable, making it highly practical for downstream tasks.

\begin{table*}[!t]
\centering
\caption{\textbf{Downstream task probing performance comparison across audio and speech benchmarks.} We evaluate models using convex gated probing (CGP), linear probing (LP), linear convex gated probing (LCGP), protobin (PB), visual query tokens (VQT), and Head2Toe (H2T). BAT outperforms other baselines and CGP significantly outperforms all probing methods. The DCASE2016 Task 2 SED reports frame-wise micro-averaged mAP and event-onset detection micro-averaged F1 score. All other tasks report the macro-averaged mAP, accuracy and F1 scores. The best model for each probing method is \textbf{highlighted}.}
\label{tab:model_comparison}
\scriptsize
\setlength\tabcolsep{0pt}
\begin{tabular*}{\textwidth}{@{\extracolsep{\fill}} ll*{12}{c}}
\toprule

& & \multicolumn{2}{c}{AS-2M} & \multicolumn{2}{c}{AS-20k} & \multicolumn{2}{c}{ESC-50} & \multicolumn{2}{c}{SC-v2} & \multicolumn{2}{c}{SED} & \multicolumn{2}{c}{HSN}\\
\cmidrule(lr){3-4} \cmidrule(lr){5-6} \cmidrule(lr){7-8} \cmidrule(lr){9-10} \cmidrule(lr){11-12} \cmidrule(lr){13-14}
\textbf{Method} & \textbf{Model} & mAP & F1 & mAP & F1 & mAP & Acc & mAP & Acc & frame-mAP & onset-F1 & mAP & F1\\
\midrule

\textcolor{gray!65}{Reported} & \textcolor{gray!65}{SSLAM} & \textcolor{gray!65}{50.2} & \textcolor{gray!65}{-} & \textcolor{gray!65}{40.9} & \textcolor{gray!65}{-} & \textcolor{gray!65}{-} & \textcolor{gray!65}{96.2} & \textcolor{gray!65}{-} & \textcolor{gray!65}{98.1} & \textcolor{gray!65}{-} & \textcolor{gray!65}{-} & \textcolor{gray!65}{-} & \textcolor{gray!65}{-} \\

\textcolor{gray!65}{Finetune} & \textcolor{gray!65}{EAT} & \textcolor{gray!65}{48.6} & \textcolor{gray!65}{-} & \textcolor{gray!65}{40.2} & \textcolor{gray!65}{-} & \textcolor{gray!65}{-} & \textcolor{gray!65}{95.9} & \textcolor{gray!65}{-} & \textcolor{gray!65}{98.3} & \textcolor{gray!65}{-} & \textcolor{gray!65}{-} & \textcolor{gray!65}{-} & \textcolor{gray!65}{-}\\

& \textcolor{gray!65}{BEATs} & \textcolor{gray!65}{48.0} & \textcolor{gray!65}{-} & \textcolor{gray!65}{38.3} & \textcolor{gray!65}{-} & \textcolor{gray!65}{-} & \textcolor{gray!65}{95.6} & \textcolor{gray!65}{-} & \textcolor{gray!65}{98.3} & \textcolor{gray!65}{-} & \textcolor{gray!65}{-} & \textcolor{gray!65}{-} & \textcolor{gray!65}{-}\\
\midrule

\multirow{4}{*}{Finetune} & \colorbox{posgreen!20}{BAT} & \textbf{48.85} & \textbf{36.62} & \textbf{41.59} & \textbf{38.20} & \textbf{98.81 $\pm$ 0.9} & \textbf{95.95 $\pm$ 1.2} & \textbf{99.80} & \textbf{99.09} & \textbf{99.71} & \textbf{98.20} & \textbf{47.05} & \textbf{34.07} \\

& SSLAM & 47.82 & 36.33 & 40.26 & 38.00 & 98.21 $\pm$ 1.0 & 94.85 $\pm$ 1.7 & 98.30 & 96.87 & 99.59 & 97.13 & 44.20 & 31.26 \\

& EAT & 47.84 & 36.38 & 40.37 & 38.03 & 98.54 $\pm$ 1.0 & 95.05 $\pm$ 1.4 & 97.50 & 96.43 & 99.70 & 97.25 & 42.32 & 31.98 \\

& BEATs & 47.02 & 36.02 & 36.73 & 32.18 & 98.01 $\pm$ 1.4 & 94.35 $\pm$ 2.3 & 99.77 & 98.85 & 99.67 & 97.34 & 18.59 & 13.00 \\
\midrule

\multirow{4}{*}{\colorbox{posgreen!20}{CGP}} & \colorbox{posgreen!20}{BAT} & {\textbf{45.03}} & {\textbf{37.94}} & {\textbf{37.70}} & {\textbf{35.22}} & \textbf{98.13 $\pm$ 1.2} & \textbf{94.55 $\pm$ 1.7} & {\textbf{99.80}} & {\textbf{98.92}} & {\textbf{99.15}} & {\textbf{94.27}} &{\textbf{41.53}} & {\textbf{29.49}} \\

& SSLAM & {42.75} & {34.22} & {34.53} & {32.06}  & {97.25 $\pm$ 1.5} & {92.55 $\pm$ 2.2} & {99.51} & {98.08} & {97.97} & {90.64} & {36.14} & {25.75} \\

& EAT   & {42.98} & {34.48} & {35.34} & {32.81} & {97.53 $\pm$ 1.2} & {93.10 $\pm$ 1.8} & {99.61} & {98.19} & {98.31} & {90.03} & {37.90} & {24.53} \\

& BEATs & {41.89} & {33.41} & {33.01} & {31.53} & {95.85 $\pm$ 0.9} & {89.15 $\pm$ 1.0} & {99.45} & {97.91} & {98.65} & {93.94} & {22.78} & {9.30}\\
\midrule

\multirow{4}{*}{PB} & \colorbox{posgreen!20}{BAT} & \textbf{42.94} & \textbf{35.94} & \textbf{35.98} & \textbf{33.88} & {\textbf{98.95 $\pm$ 0.7}} & {\textbf{95.75 $\pm$ 0.7}} & \textbf{99.76} & \textbf{98.68} & \textbf{98.03} & \textbf{91.52} & \textbf{36.14} & \textbf{26.28} \\

& SSLAM & 38.53 & 31.57 & 32.01 & 30.50 & 96.38 $\pm$ 1.3 & 90.70 $\pm$ 1.4 & 98.04 & 95.84 & 93.87 & 80.34 & 28.20 & 15.52 \\

& EAT & 39.08 & 31.81 & 32.87 & 31.11 & 96.83 $\pm$ 1.1 & 91.45 $\pm$ 1.7 & 98.52 & 96.50 & 95.15 & 84.77 & 26.47 & 18.63 \\

& BEATs & 37.73 & 30.42 & 31.62 & 28.43 & 95.31 $\pm$ 1.4 & 88.75 $\pm$ 1.8 & 99.27 & 97.47 & 97.95 & 90.48 & 17.32 & 3.85\\
\midrule

\multirow{4}{*}{VQT} & \colorbox{posgreen!20}{BAT} & \textbf{40.45} & \textbf{29.88} &  \textbf{34.23} & \textbf{34.47} & \textbf{94.88 $\pm$ 1.8} & \textbf{91.50 $\pm$ 2.1} & \textbf{99.21} & \textbf{97.65} & N/A & N/A & \textbf{23.61} & \textbf{14.08} \\

& SSLAM & 37.78 & 28.55 & 31.21 & 31.34 & 91.51 $\pm$ 1.6 & 88.70 $\pm$ 1.6 & 97.80 & 95.25 & N/A & N/A & 12.14 & 5.57 \\

& EAT & 39.11 & 28.55 & 31.89 & 31.98 & 92.48 $\pm$ 2.6 & 89.50 $\pm$ 2.2 & 98.32 & 96.16 & N/A & N/A & 12.97 & 3.78 \\

& BEATs & 37.90 & 26.50 & 31.56 & 30.32 & 86.21 $\pm$ 1.9 & 82.05 $\pm$ 1.5 & 98.45 & 96.38 & N/A & N/A & 8.24 & 5.29\\
\midrule

\multirow{4}{*}{H2T} & \colorbox{posgreen!20}{BAT} & \textbf{38.28} & \textbf{31.60} & \textbf{29.59} & \textbf{30.43} & \textbf{97.64 $\pm$ 1.1} & \textbf{92.45 $\pm$ 1.5} & \textbf{94.62} & \textbf{92.02} & \textbf{94.32} & \textbf{80.86} & \textbf{14.67} & \textbf{4.57} \\

& SSLAM & 37.28 & 30.79 & 27.41 & 29.96 & 95.57 $\pm$ 1.2 & 89.65 $\pm$ 1.6 & 87.36 & 86.62 & 88.74 & 66.89 & 9.10 & 1.17 \\

& EAT & \textbf{38.28} & 31.56 & 28.35 & 30.97 & 96.54 $\pm$ 1.1 & 91.35 $\pm$ 1.1 & 93.84 & 91.43 & 90.09 & 69.90 & 9.82 & 3.67\\

& BEATs & 36.55 & 31.10 & 25.56 & 29.12 & 94.27 $\pm$ 1.6 & 86.90 $\pm$ 1.9 & 95.57 & 92.81 & 92.94 & 80.27 & 4.06 & 2.54\\
\midrule

\multirow{4}{*}{LCGP} & \colorbox{posgreen!20}{BAT} & \textbf{35.99} & \textbf{26.89} & \textbf{29.95} & 26.15 & \textbf{97.52 $\pm$ 1.1} & \textbf{92.25 $\pm$ 1.4} & \textbf{91.41} & \textbf{89.26} & \textbf{95.73} & \textbf{83.30} & \textbf{13.47} & 7.52 \\

& SSLAM & 33.92 & 24.53 & 28.15 & 24.06 & 95.48 $\pm$ 1.4 & 89.65 $\pm$ 1.8 & 79.20 & 80.63 & 90.40 & 73.25 & 12.83 & 3.33 \\

& EAT & 35.42 & 25.88 & 28.84 & \textbf{26.74} & 96.75 $\pm$ 0.9 & 91.45 $\pm$ 1.8 & 90.09 & 88.70 & 92.41 & 76.03 & 12.08 & \textbf{8.65} \\

& BEATs & 33.72 & 23.33 & 27.91 & 26.13 & 93.87 $\pm$ 1.1 & 86.40 $\pm$ 1.5 & 93.70 & 91.22 & 95.10 & 83.15 & 5.25 & 4.69\\
\midrule

\multirow{4}{*}{LP} & \colorbox{posgreen!20}{BAT} & 31.19 & 20.21 & 26.15 & 22.41  & \textbf{95.25 $\pm$ 1.3} & \textbf{89.20 $\pm$ 2.1} & 75.74 & 78.12 & 94.21 & \textbf{82.38} & 9.21 & \textbf{5.28} \\

& SSLAM & 26.96 & 17.70 & 21.14 & 17.61 & 94.10 $\pm$ 0.9 & 87.65 $\pm$ 1.9 & 56.28 & 70.22 & 86.33 & 66.09 & 8.52 & 3.95 \\

& EAT & 27.30 & 18.21 & 21.60 & 16.98 & 91.07 $\pm$ 0.7 & 84.75 $\pm$ 1.8 & 74.31 & 78.03 & 88.67 & 70.21 & \textbf{10.77} & 2.75 \\

& BEATs & \textbf{31.40} & \textbf{22.05} & \textbf{26.26} & \textbf{23.17} & 93.07 $\pm$ 1.1 & 86.50 $\pm$ 1.6 & \textbf{92.60} & \textbf{90.44} & \textbf{94.61} & 81.72 & 5.33 & 4.79\\

\bottomrule
\end{tabular*}
\end{table*}

\section{Benchmark Results}
\autoref{tab:model_comparison} reports the main benchmarking results of our work.

\textbf{Setup.} We validate the models across AS-20k, AS-2M, ESC-50~\citep{piczak2015esc50}, Speech Commands V2 (SC-v2)~\citep{warden2018_speechcommands}, Sound Event Detection (SED)~\citep{Mesaros2018_TASLP}, out-of-distribution classification using the High Sierra Nevada (HSN) task in BirdSet~\citep{rauch2025_birdset}, and Automatic Speech Recognition (ASR) using LibriSpeech~\citep{panayotov2015librispeech}. The datasets are detailed in \autoref{app:datasets}. In \autoref{tab:topaudiossl}, we reproduce EAT and SSLAM by replicating their finetuning using their published pretrained weights. This final evaluation uses slightly more refined hyperparameters for BAT, EAT, SSLAM, and BEATs. BEATs is included as a discrete-target masked prediction baseline, complementing EAT and SSLAM. Comprehensive hyperparameter details for all protocols are provided in \autoref{app:hyperparams}.

\textbf{Frozen-feature probing.} BAT outperforms prior models for nearly all probing methods, while simple LP remains less reliable as an indicator for representation quality. Results for PB demonstrate that the last-layer embeddings from BAT contain richer information than prior models, including discrete-token masked prediction models models such as BEATs. The strong performance of CGP, even under severe domain shift (HSN) or for dense tasks (SED), demonstrates its faithfulness in assessing SSL embeddings without the hurdles of finetuning. This boosts confidence in adopting CGP as the default evaluation paradigm in future audio SSL research, providing a more transparent line of progress. Moreover, we allow competing probing methods an advantage to strengthen this argument. A faithful frozen probing should be a direct reflection of raw SSL embeddings. However, we find that all non-prototype probing methods, even VQT, perform worse without a learnable layer norm applied to the embeddings. Additionally, we find that Lasso regularization in H2T is highly sensitive and requires careful tuning. For AudioSet, Lasso regularization prevents learning, and we set its weight to zero. For other datasets, a very small Lasso weight of $1\mathrm{e}{-4}$ achieves reasonable performance. For VQT, we tune the number of learnable query tokens, and 10 yields the best performance.

\textbf{Reproducibility and finetuning.} We observe a notable discrepancy between performance reported in the literature and our reproductions, particularly for SSLAM and AS-2M. Although EAT results are slightly closer, we also cannot reproduce their pretrained SSL model using their exact recipe (compare row 1 of \autoref{tab:input-ablation} with CGP probing using EAT's published weights in \autoref{tab:topaudiossl}). However, BAT surpasses the reported AS-20k results and several other benchmarks, indicating that the lack of reproducibility for AS-2M is not solely due to suboptimal hyperparameter tuning. Instead, our investigations attribute this to the AS-2M sampling procedure during training, a legacy from SSAST~\citep{gong2022ssast}. Hence, we document the training sampler transparently in our code base. As shown in \autoref{fig:cgp_layerwise_info}, the location of task-relevant latent information shifts significantly in our model compared to baselines, which likely affects suitable hyperparameters. \autoref{app:cgp_blocks} shows the CGP gating weights for all datasets and models.

\textbf{Task-relevant latent information.} As illustrated in \autoref{fig:cgp_layerwise_info} and \autoref{app:cgp_blocks}, a major factor for the superiority of BAT is an increase in the utilized capacity of the encoder by offloading the reconstruction task to an expressive decoder. This is more intuitive to see for  MAE~\citep{he2022_maskedauto} than for MLR. However, we caution against overgeneralizing "later-is-better" as a universal metric. Different architectures or SSL objectives may yield encoders that peak in the middle layers while maintaining competitive performance. Therefore, the semantic shift observed in BAT may indicate improvement compared to similar MLR models, but we have no evidence that this is sufficient to claim that, if the best features are not in the final layers, the model as a whole is underutilized. This makes CGP a well-suited post-hoc evaluation probe for addressing such concerns in SSL. 

\textbf{Automatic Speech Recognition.}
We use a two-layer LSTM on top of the frozen backbone, followed by a layer norm and the linear classifier. Speech models typically rely on fine-grained temporal embeddings, and our most relevant prior work does not report ASR results. The embeddings of our baselines have a temporal resolution of 160~ms. Thus, we add a learnable transposed convolution layer before the LSTMs to upsample the temporal patch tokens by a factor of 8, achieving 20 ms temporal resolution. The models are trained on LibriSpeech (100 hrs clean). \autoref{tab:asr} reports Word Error Rate (WER) and Character Error Rate (CER). 


\begin{table}[H] 
    \centering
    \small
    \caption{\textbf{Automatic speech recognition.} Probing on the test set.}
    \label{tabapp:asr}
    \begin{tabular}{@{}lcccc@{}}
    \toprule
    \textbf{Metrics} & \colorbox{posgreen!20}{BAT} & SSLAM & EAT & BEATs \\
    \midrule
    CER & \textbf{7.27} & 10.06 & 9.26 & 8.12 \\
    WER & \textbf{22.18} & 29.05 & 27.14 & 24.67 \\
    \bottomrule
    \label{tab:asr}
    \end{tabular}
\end{table}

\section{Conclusion}
This work revisited the training and evaluation practices in audio self-supervised learning (SSL). It showed that the reliance on finetuning for state-of-the-art (SOTA) AudioSet results has made progress less transparent, since improvements become harder to disentangle from dataset-specific tuning and reproducibility issues. To alleviate these challenges, we proposed the prototype-based Convex Gated Probing (CGP). By efficiently aggregating features across all layers, CGP significantly closes the performance gap between frozen evaluation and finetuning, outperforming the prior SOTA probing method, Protobin. Guided by CGP as a fast and reliable post-hoc evaluation probe, we introduced the Better Audio Transformer (BAT), a fully modernized implementation of masked latent regression for audio SSL. BAT incorporates the sigmoid-gating mechanism for self-attention, a recent advancement in LLMs, which not only improves the model overall but also enables us to drop the Data2Vec MLP-target heuristic and generate better targets for SSL by using each ViT block as a coherent end-to-end function. BAT also enhances representation learning by using a more expressive decoder, which offloads reconstruction from the encoder, enabling better utilization of its capacity to learn transferable features. CGP's gating weights demonstrate this shift by revealing the location of task-relevant latent information. Additionally, we refined the legacy audio preprocessing pipeline, i.e., the audio frontend. This improved frontend produces higher-quality spectrograms, enables fast batch transformations, and, most notably, alleviates the need to tune global normalization statistics for downstream deployment through local min-max normalization. BAT established new SOTA results on audio benchmarks while ensuring reproducibility. The BAT code is open-source, and we provide an implementation and evaluation pipeline for recent SOTA models used in this work.

\clearpage \newpage
\section*{Impact Statement}
This work advances the field of self-supervised audio representation learning. By improving the effective depth of encoders and introducing efficient probing methods such as CGP, we contribute to more resource-efficient model evaluation. This has the potential to positively impact society by reducing the computational overhead and carbon footprint associated with developing state-of-the-art audio models. Furthermore, enhancing the robustness of frozen embeddings could broaden access to high-performance models for researchers with limited compute resources. There are no specific negative ethical consequences unique to this work.

\section*{Author Contribution}
\textbf{Houtan Ghaffari}: Conceptualization (lead), Methodology, Implementation (lead), Formal Analysis, Validation, Writing, Visualization, Resources.
\textbf{Lukas Rauch}: Conceptualization, Methodology, Implementation, Formal Analysis, Validation, Writing, Visualization, Resources.
\textbf{Christoph Scholz}: Resources.
\textbf{Paul Devos}: Resources, Supervision.

\section*{Acknowledgements}
This research was partly funded by the Ghent University Special Research Fund (grant BOF/STA/202102/005), and partly under the BioDroneAI project (FKZ 02WDG1758D), funded by the German Federal Ministry of Research, Technology and Space (BMFTR).

\bibliography{references}
\bibliographystyle{icml2026}

\newpage
\appendix
\onecolumn

\section{Hyperparameters}
\label{app:hyperparams}

\begin{table*}[h]
\centering
\caption{\textbf{Hyperparameter configurations.} We show our hyperparameters for pretraining, finetuning, and probing. All probing methods used the same hyperparameters, with no special advantage for CGP. We tuned the number of tokens for VQT and the best Lasso for Head2Toe. Sound Event Detection (SED) refers to DCASE2016 Task 2. Automatic Speech Recognition (ASR) uses LibriSpeech. The loss function acronyms are: Binary Cross-Entropy (BCE), Asymmetric Binary Cross-Entropy (A-BCE)~\citep{Ridnik2021_asymmetricloss}, and Connectionist Temporal Classification (CTC).}
\label{tab:hyperparameters}
\resizebox{\textwidth}{!}{
\begin{tabular}{l|c|cccccc|ccccccc}
\toprule

\multirow{2}{*}{Hyperparameters} & Pretraining & \multicolumn{6}{c|}{Finetuning} & \multicolumn{7}{c}{Probing} \\  

 & AS-2M & AS-2M & AS-20K & ESC-50 & SC-v2 & HSN & SED & AS-2M & AS-20K & ESC-50 & SC-v2 & HSN & SED & ASR \\
\midrule

Optimizer & \multicolumn{14}{c}{AdamW} \\
Weight Decay & \multicolumn{14}{c}{0.05} \\
Optimizer Momentum ($\beta_1$, $\beta_2$) & \multicolumn{7}{c |}{(0.9, 0.95)} & \multicolumn{7}{c}{(0.9, 0.999)} \\

Learning Rate Scheduler & \multicolumn{14}{c}{Cosine } \\

Peak Learning Rate & 5e-4 & \multicolumn{6}{| c}{5e-5} & \multicolumn{7}{| c}{1e-3} \\

Minimum Learning Rate & \multicolumn{14}{c}{1e-6} \\

Layer-Wise Learning Rate Decay & N/A & \multicolumn{6}{| c}{0.75} & \multicolumn{7}{| c}{N/A} \\

Optimization Steps (k)         & 400 & 200 & 40 & 4 & 10 & 10 & 4 & 200 & 20 & 4  & 10 & 10 & 4 & 20\\

Learning Rate Warmup Steps (k) & 2 & 20 & 4 & 0.4 & 1 & 1 & 0.4 & 1 & 0.5 & 0.4  & 1 & 1 & 0.4 & 0.5\\

Batch Size per GPU & 12 & 96 & 48 & 48 & 256 & 64 & 4 & 96 & 48 & 48 & 256 & 64 & 4 & 64\\

GPUs & 4 & \multicolumn{13}{c}{1} \\

Masked Views & 16 & \multicolumn{13}{c}{N/A} \\

Drop path & 0.0 & \multicolumn{6}{|c}{0.1} & \multicolumn{7}{|c}{0} \\

Class-Weighted Train Sampling & False &  True (200k) & False & False & False & True (5.5k) & False & True (200k) & False & False & False & True (5.5k) & False & False\\

Mixup Chance & N/A & 0.8 & 0.9 & 0.9 & 0.9 & 0.9 & N/A & \multicolumn{7}{|c}{N/A} \\

Mixup Beta & N/A & 0.8 & 0.8 & 0.8 & 0.8 & 0.8 & N/A & \multicolumn{7}{|c}{N/A} \\

Color Noise Chance & N/A & 0 & 0 & 0.3 & 0.3 & 0.3 & 0.3 & \multicolumn{7}{|c}{N/A} \\

SpecAug Frame Masking (Time, Freq)  & N/A & (64,16) & (64,16) & (32, 16) & (16, 8) & (64, 16) & (64, 16) & \multicolumn{7}{|c}{N/A} \\

Loss Function & MSE & BCE & A-BCE & BCE & BCE & BCE & BCE & A-BCE & A-BCE & BCE & BCE & BCE & BCE & CTC\\

Prototypes (CGP \& Protobin) & \multicolumn{7}{c}{N/A} & \multicolumn{6}{|c}{10,000} & N/A \\

Visual Query Tokens (only VQT) & \multicolumn{7}{c}{N/A} & \multicolumn{6}{|c}{10} & N/A \\

Lasso Regularization (only H2T) & \multicolumn{7}{c}{N/A} & 0 & 0 & 0.0001 & 0.0001 & 0.0001 & 0.0001 & N/A \\
\bottomrule
\end{tabular}
}
\end{table*}

\section{Datasets}
\label{app:datasets}
\autoref{tab:datasets} provides an overview of these datasets. The pretraining is done solely on AS-2M.

\begin{table}[H] 
\centering
\small
\caption{\textbf{Overview of datasets.}}
\label{tabapp:dataset}
\begin{tabular}{@{}lccccc@{}}
\toprule
\textbf{Dataset} & \textbf{Train} & \textbf{Validation} & \textbf{Test} & \textbf{\#Classes} & \textbf{Segments [seconds]}\\ \midrule
\texttt{AudioSet unbalanced (AS-2M)}  & $1,912,024$ & -       & $18,886$ & $527$ & 10 \\
\texttt{AudioSet balanced (AS-20k)} & $20,550$    & -       & $18,886$ & $527$ & 10 \\
\texttt{Environmental Sound Classification (ESC-50)}  & $1,600$     & -       & $400$    & $50$  & 5  \\ 
\texttt{Speech Commands v2 (SC-v2)}  & $84,848$    & $9,982$ & $4,890$  & $12$  & 1  \\
\texttt{High Sierra Nevada (HSN)}         & $5,460$    & - & $12,000$  & $21$  & 5 (random crop)  \\
\texttt{DCASE 2016 Task 2 (SED)}    & $44$    & $14$ & $14$  & $11$  & 120  \\
\texttt{LibriSpeech (ASR)}          & $28,539$    & $2703$ & $2620$  & $29$  & [1, 25] (variable)  \\


 \bottomrule
 \label{tab:datasets}
\end{tabular}
\end{table}
\clearpage
\newpage
\section{Additional Results}
\autoref{tab:vits_appendix_as} reports the probing and finetuning results for a pretrained BAT using the ViT-small architecture. All hyperparameters for both SSL and downstream experiments are the same as those of the ViT-base model.

\begin{table}[hp]
\centering
\caption{\textbf{ViT-Small downstream adaptation on AudioSet.}}  
\label{tab:vits_appendix_as}
\scriptsize
\setlength\tabcolsep{4pt}
\begin{tabular*}{0.62\textwidth}{@{\extracolsep{\fill}}llcccc}
\toprule
& & \multicolumn{2}{c}{AS-2M} & \multicolumn{2}{c}{AS-20K} \\
\cmidrule(lr){3-4} \cmidrule(lr){5-6}
\textbf{Method} & \textbf{Model} 
& mAP & F1 
& mAP & F1 \\
\midrule

Finetune 
& \colorbox{posgreen!20}{BAT-S}
& \textbf{45.77} 
& \underline{30.86}
& \textbf{37.55} 
& \textbf{34.38} \\

\colorbox{posgreen!20}{CGP}
& \colorbox{posgreen!20}{BAT-S}
& \underline{41.63} 
& \textbf{34.43} 
& \underline{33.68} 
& \underline{31.70} \\

PB 
& \colorbox{posgreen!20}{BAT-S}
& 40.19 
& 32.26 
& 32.63 
& 29.27 \\

VQT 
& \colorbox{posgreen!20}{BAT-S}
& 40.01 
& 33.28 
& 30.76 
& 29.90 \\

H2T 
& \colorbox{posgreen!20}{BAT-S}
& 32.74 
& 23.78
& 25.03 
& 26.37 \\

LCGP 
& \colorbox{posgreen!20}{BAT-S}
& 27.53 
& 15.77 
& 21.56 
& 11.81 \\

LP 
& \colorbox{posgreen!20}{BAT-S}
& 25.40 
& 12.97
& 21.26 
& 11.45 \\

\bottomrule
\end{tabular*}
\end{table}

\section{Probing Methods Computational Costs and Latency}\label{app:cost_latency}
\autoref{tab:performance_comparison} reports the forward pass computation cost, GPU RAM consumption, and latency. These results were obtained with PyTorch 2.8.0, CUDA 12.8, and an NVIDIA RTX A6000 (48 GB). The last column is the Multiply-Accumulate (MAC) operations. Note that we use the cls-tokens of all layers for the Head2Toe. We find that using all features of the model is not only computationally prohibitive (a linear classifier using all the features from a ViT-base model requires 2.42 billion parameters) but also detrimental to performance, making training difficult to converge. Furthermore, we find that Lasso regularization rarely has a positive effect, prevents convergence on the AudioSet task, and does not easily yield sparse weights on audio datasets. Note that the VQT method requires the full ViT model during probing, and we cannot train it without the model on pre-extracted features. 

\begin{table}[htbp]
    \centering
    \scriptsize
    \caption{\textbf{Computational cost and latency of the probing methods.}}
    \label{tab:performance_comparison}
    \begin{tabular}{llcccccc}
        \toprule
        \textbf{Method} & \textbf{Prototypes / Queries} & \textbf{Classes} & \textbf{Params (M)} & \textbf{Input \& Shapes} & \textbf{Peak VRAM (MB)} & \textbf{Latency (ms)} & \textbf{MAC (G)} \\
        \midrule
        Linear        & N/A   & 527 & 0.41  & last layer cls=(1, 768) & 28.72  & $0.09 \pm 0.02$ & 0.0004 \\
        
        Linear-CGP    & N/A   & 527 & 0.41  & cls=(1, 12, 768) & 28.72  & $0.14 \pm 0.02$ & 0.0004 \\
        
        Head2Toe & N/A   & 527 & 4.88  & cls=(1, 12, 768) & 63.84  & $0.10 \pm 0.02$ & 0.0050 \\
        
        VQT      & 10    & 527 & 49.07 & spectrogram=(1, 1024, 128)      & 637.96 & $10.30 \pm 0.25$ & 53.2681 \\
        
        Protobin & 10000 & 527 & 12.95 & last layer patch=(1, 768, 64, 8)     & 166.27 & $0.91 \pm 0.02$ & 3.9374 \\
        
        CGP      & 10000 & 527 & 23.49 & [patch=(1, 12, 768, 64, 8), cls=(1, 12, 768)] & 184.20 & $0.72 \pm 0.03$ & 3.9556 \\
        \bottomrule
    \end{tabular}
\end{table}

\section{Resources}
Most of the experiments were computed on NVIDIA A100 GPUs. This includes the multi-GPU SSL runs for pretraining and some of the downstream experiments. Some unit tests and ablations during development, as well as some downstream experiments, were conducted on NVIDIA RTX A6000 and NVIDIA RTX 4090 GPUs. We could reproduce our results, to within negligible decimal places, across multiple devices and library versions. 

\newpage
\section{CGP Block Weights}\label{app:cgp_blocks}

\begin{figure}[h!]
\centering
\includegraphics[width=1.0\linewidth]{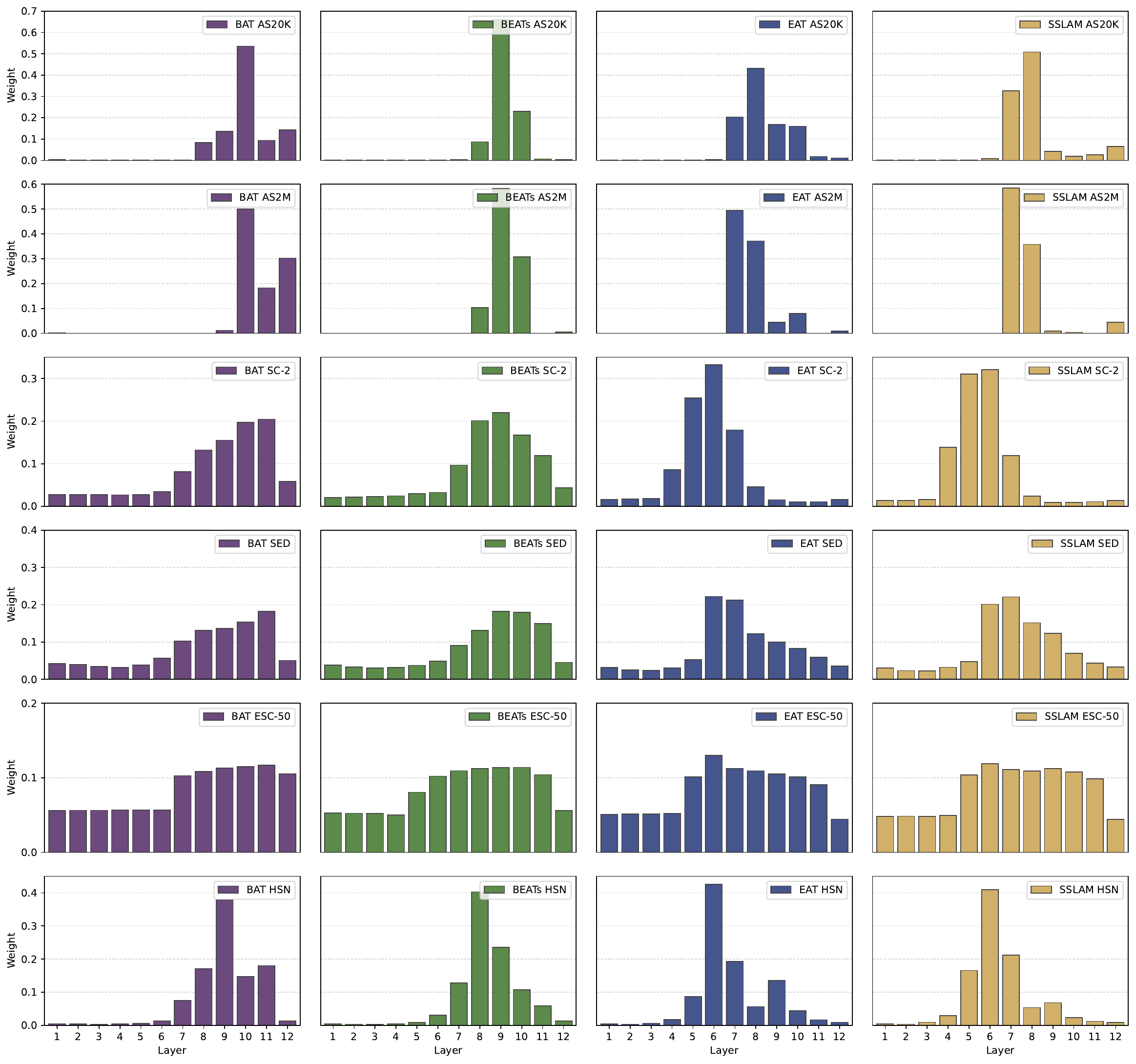}
\caption{\textbf{CGP layer-wise gating weights across datasets.} The task-relevant information in BAT is pushed towards the later blocks even more than BEATs, which is a contrastive method. The difference in the distribution of the gating weights in this figure and those in \autoref{fig:cgp_layerwise_info} of the main paper is due to the final benchmarking experiments for AudioSet leveraging the asymmetric loss. We plot these CGP models here. The ablation in the methodological exposition used the conventional cross-entropy loss to avoid additional factors affecting the findings due to the asymmetric loss's sensitivity to its hyperparameters.}
\label{fig:cgp_attention_esc_sc}
\end{figure}

\end{document}